\documentclass[onecolumn,authoryear]{els-mrw} 

\usepackage{amsmath,amssymb,amsfonts,amsthm,makeidx,graphicx,natbib}
\usepackage{ogonek}
\usepackage{txfonts}
\usepackage{helvet}

\begin{document}

\chapter{White Dwarf Variability}\label{chap1}

\author[1]{Keaton J.\ Bell}

\address[1]{\orgname{CUNY Queens College}, \orgdiv{Department of Physics}, \orgaddress{65-30 Kissena Blvd, Flushing, NY, 11367, USA}}


\maketitle


\begin{abstract}[Abstract]
There are a few different mechanisms that can cause white dwarf stars to vary in brightness, providing opportunities to probe the physics, structures, and formation of these compact stellar remnants. The observational characteristics of the three most common types of white dwarf variability are summarized: stellar pulsations, rotation, and ellipsoidal variations from tidal distortion in binary systems. Stellar pulsations are emphasized as the most complex type of variability, which also has the greatest potential to reveal the conditions of white dwarf interiors.
\end{abstract}

\begin{glossary}[Key Points]
\term{} Variations in the brightness of white dwarf stars can reveal details about the physical conditions of these compact stellar remnants:
\begin{itemize}
\item Stellar pulsations cause some white dwarfs to vary in brightness at frequencies that are resonant within the star, sensitively probing the conditions of the stellar interior.
\item Rotation of white dwarfs can cause brightness variations if the surface is not uniformly bright, enabling the measurement of rotation rates.
\item Tidal distortion of white dwarfs in tight binaries can cause ellipsoidal variations when the projected area of the white dwarf varies through the orbit, which can constrain the inclination angle of the system.
\end{itemize}

\term{} Variations from stellar pulsations have the most complicated characteristics:
\begin{itemize}
\item A white dwarf pulsates when passing through an instability strip, where the temperature is right given its atmospheric composition for pulsations to be globally driven.
\item Not all resonant frequencies are driven to observable amplitude, and the pulsation modes that we observe vary from star to star.
\item The pulsations of white dwarfs are non-radial gravity-mode pulsations with buoyancy as a restoring force. On average these pulsations are evenly spaced in period, with variations from even spacing that encode information about interior structure.
\item Rotation splits pulsations into multiplets of frequencies that can be represented by different spherical harmonic brightness variations.
\item Interactions of pulsation modes with each other or the convection zone can affect the pulse shape, affect coherence of pulsations, or cause sudden brightness increases as pulsation energy is released as pulsational outbursts.
\end{itemize}

\end{glossary}

\begin{glossary}[Glossary and Nomenclature]
\term{White dwarf stars} are the compact objects that remain after most stars exhaust all of their nuclear fuel.

\term{Light curves} consist of measurements of stellar brightness recorded at different times.

\term{Periodograms} can be computed from light curves to detect and characterize frequencies of brightness variations.

\term{Gravity mode pulsations} are stellar oscillations with buoyancy as the restoring force.

\term{Instability strips} are atmosphere-dependent combinations of masses and effective (surface) temperatures where pulsations are driven.

\term{DAVs} (ZZ Ceti variables) are hydrogen-atmosphere white dwarfs that pulsate with effective temperatures around 10,500--13,000\,K.

\term{DBVs} (V777 Her variables) are helium-atmosphere white dwarfs that pulsate with effective temperatures around 22,000--31,000\,K.

\term{GW Vir} stars are pulsating hot white dwarfs or pre-white dwarfs with effective temperatures in the range 80,000--180,000\,K.

\term{Asteroseismology} is the collection of techniques that are used to interpret pulsation properties to infer stellar parameters.

\term{Ellipsoidal variations} are brightness variations caused by tidal distortions of stars in tight binary systems.
\end{glossary}

\section{Introduction}\label{chap1:sec1}
\setlength{\parindent}{12pt}
Most white dwarf stars are remarkably constant in brightness over a wide range of timescales \citep{2017MNRAS.468.1946H}. Being photometrically stable makes white dwarfs valuable as flux standards for calibrating (spectro)photometric systems, enabling flux measurements to be converted to physical units. The relative simplicity of modeling pure-hydrogen atmospheres motivates the use of white dwarf stars as calibrators for observations made with the Hubble Space Telescope \citep{2014AJ....147..127B}, Spitzer \citep{2011AJ....141..173B}, and JWST \citep{2022AJ....163..136M}.

However, some white dwarfs make exceptions and vary in brightness in ways that reveal more about their physical properties. Observations that constrain the properties of these objects are astrophysically valuable for multiple reasons, including:
\begin{enumerate}
    \item White dwarf stars are the final evolutionary state for 97\% of stars in the Milky Way \citep{2008ApJ...676..594K}. They serve as observational boundary conditions for shaping our understanding of stellar evolution and its effect on planetary systems, especially during the rapid late stages when considerable mass loss occurs. 
    \item White dwarfs are compact objects, providing remote laboratories for studying the behavior of matter under extreme physical conditions \citep{2022PhR...988....1S,2022FrASS...9....6I}. 
    \item White dwarfs evolve by cooling, and their temperatures or luminosities can be used to constrain ages of stellar populations (cosmochronology; \citealt{1987ApJ...315L..77W,2001PASP..113..409F}). They are the progenitors of Type Ia supernovae, which provide a rung in the cosmological distance ladder and reveal the dark energy content of the Universe \citep{1998AJ....116.1009R}. 
\end{enumerate}
The quality of astrophysical insights afforded by white dwarfs are limited by our understanding of their structures.

This article is concerned with the most common causes of photometric variability: stellar pulsation (section \ref{sec:pulsations}), rotation (\ref{sec:rotation}), and ellipsoidal variations in binaries (\ref{sec:EVs}). Most of the chapter will discuss pulsations, as this behavior is the most rich and complicated. The characteristics of white dwarf variability are summarized, with references to how we interpret these observations with basic theory. The chapter does not aim to provide a thorough review of the scientific literature, but the references given provide some reasonable entry points. The information in this chapter should be useful for characterizing the type of variability recorded in a white dwarf light curve, as a starting point for more detailed investigation.

\subsection{Light curve data and periodogram analysis}\label{sec:periodogram}

This chapter is focused on the causes of brightness variations of white dwarfs, which is really only of interest once we know how to answer the question: is this white dwarf varying in brightness? And if so, what are the characteristics of these variations? The type of data that can answer this question is time series photometry: repeated measurements of the brightness of the star from images collected at different times that constitute a light curve. A light curve includes times of observation, measured flux (often normalized to have a mean of one to represent relative flux), and ideally uncertainties on the flux measurements. 

Astronomers have been collecting increasingly more time series observations as telescopes aim to monitor the sky for variability on various timescales.  Ground-based surveys are increasingly operating in the time domain, such as ASAS-SN \citep{ASAS-SN} and ZTF \citep{ZTF}, and soon the LSST survey from the Vera C.\ Rubin Observatory \citep{LSST}, which will extend such studies to much fainter stars. Space missions that were primarily motivated by exoplanet science, such as \textit{Kepler} \citep{Kepler}, \textit{K2} \citep{K2}, and TESS \citep{TESS}, have provided precise, continuous light curves with coverage spanning from months to years. The continuous nature of these light curves limits confusion over the timescales of variability that occurs when there are gaps in the data. The value of continuous observations of white dwarf stars motivated the Whole Earth Telescope \citep{WET} prior to the space photometry era: a collaboration between observatories distributed in longitude that can hand off observations around the globe to maintain coverage despite limited target visibility at each observing site.

With the right data in hand, the question becomes: is this light curve consistent with constant brightness, or has it recorded a significant variation in brightness? The statistical technique used to answer this question most often is periodogram analysis. Based on the Fourier Transform, various periodogram methods transform time domain photometric data into the frequency domain, revealing how much the light curve varies at different frequencies. This transformation decomposes the light curve into a set of sinusoidal basis functions, each with a different frequency. Plotted for a series of tested frequencies (or sometimes their reciprocal: periods), periodograms show either the best-fit amplitude of a sinusoid at each sampled frequency, or the power, which is the square of the amplitude. 
Examples of periodograms for five pulsating hydrogen-atmosphere white dwarfs are displayed in the bottom panels of Fig.~\ref{fig:davstrip}, where amplitude of variability is displayed in percent, and frequency is given in microHertz (1\,$\mu$Hz = $10^{-6}$\,Hz = $10^{-6}$\,second$^{-1}$). Fourier methods can reveal signals that are not immediately apparent when visually inspecting a light curve, because the signal amplitudes may be smaller than the per-point measurement noise. In the periodogram, measurement noise gets distributed across all frequencies, while brightness variations appear as peaks that rise significantly above this noise floor at particular frequencies, which can be identified with statistical methods \citep{2018ApJS..236...16V}.

Accurate period measurements are most reliably determined from continuous light curves that are rapidly sampled. Gaps in data and low-frequency sampling (such that signals are sampled fewer than twice per period, i.e., above the Nyquist frequency) introduce signal aliases to periodograms \citep{2018ApJS..236...16V}, like reflections of the true signals at inaccurate frequencies. This article will focus on high-cadence, nearly continuous data from space where these issues are largely avoided.

\section{White dwarf pulsations}\label{sec:pulsations}

Some white dwarf stars pulsate. If their conditions are just right, they will oscillate globally at one or more natural frequencies that are resonant within the star. Pulsation periods are typically in the range 2--20 minutes, with relative amplitudes of photometric variations observed to range from parts per thousand up to ten percent \citep{2006ApJ...640..956M}. The pulsations compress the stellar material, causing temperature and corresponding brightness variations at the photosphere. Resonant frequencies of the star can be measured from light curves via periodogram analysis (\ref{sec:periodogram}). This enables these white dwarfs to be studied with the powerful techniques of asteroseismology: the analysis of resonant frequencies to infer properties of the stellar interior. The pulsations propagate through the white dwarf, and their frequencies are sensitively tuned by the precise structure of the star. Asteroseismic interpretation of pulsation signals has the potential to probe white dwarf interior structures and physics, which have important astrophysical consequences. The most common approach to inferring interior structures of white dwarfs is by forward modeling, whereby pulsation periods calculated from numerical models of white dwarfs with different structures are fit to the observed pulsation periods \citep[e.g.,][]{2022FrASS...9.9045G}. 
We discuss the main classes of pulsating white dwarf, with a focus on the most populous ``DAV'' class where observational trends have been most clearly established.

\subsection{Driving of pulsations in the instability strips}

A small subset of white dwarf stars are pulsating---those with just the right conditions that oscillations are self-driven and amplified by the stellar interior. White dwarfs that meet these conditions for pulsations are within pulsational ``instability strips''---regions of observational parameter space where white dwarfs are observed to pulsate. There are three primary instability strips for white dwarfs, and white dwarfs that pulsate in each of these regimes are classified as either DAVs, DBVs, or GW Vir stars. The main classes of pulsating white dwarfs and their respective driving mechanisms are described below.

The driving mechanism is responsible for exciting some vibrational modes of a star to observable amplitude. The driving mechanisms operating in white dwarfs are not perfectly understood; it is unknown why only select pulsation modes are excited in individual stars, while other modes with similar characteristics are not detected \citep[e.g.,][]{2013MNRAS.430...50C}. Fortunately, the specifics of the driving mechanism do not have much bearing on the interpretation of detected pulsation frequencies. 
However the modes are driven, they reveal some of the resonant frequencies of the pulsating stars, which can be studied to infer details of their interior structures.

\begin{figure}[t]
\centering
\includegraphics[width=.68\textwidth]{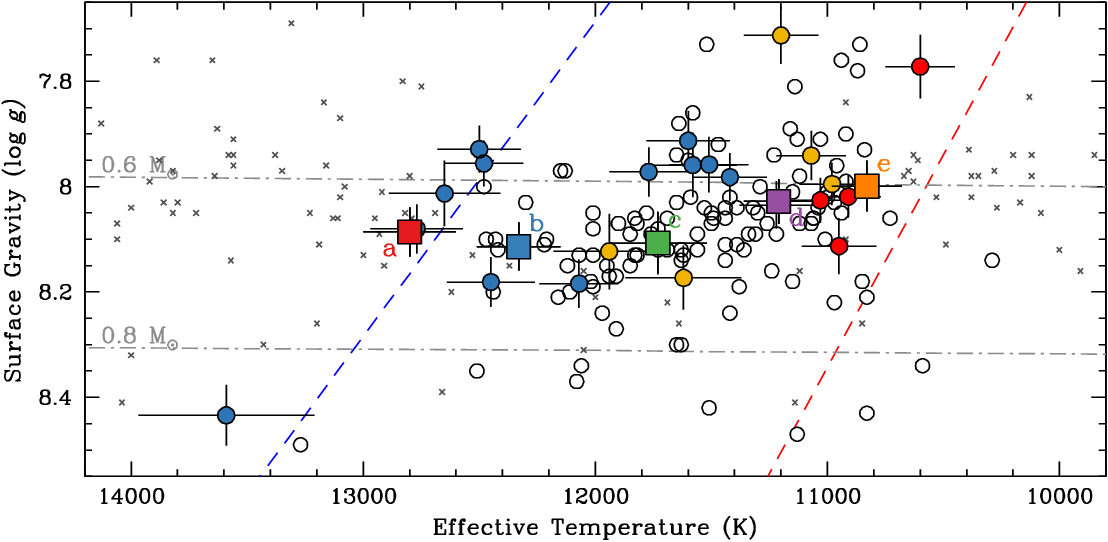}
\includegraphics[width=.68\textwidth]{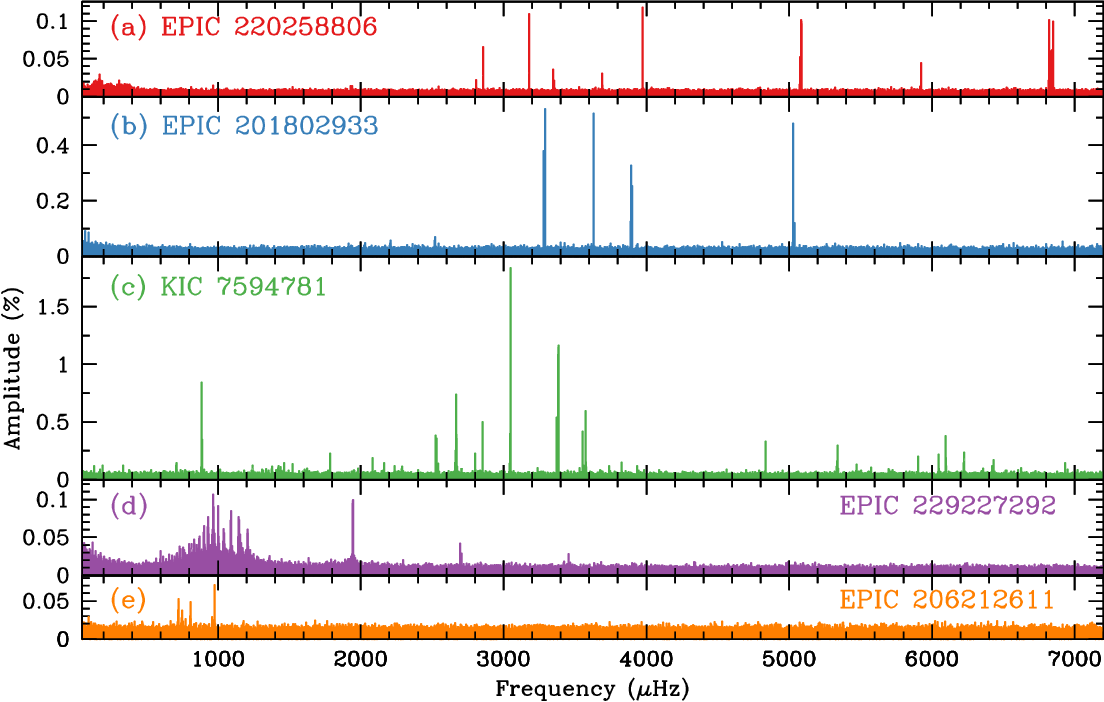}
\caption{Examples of DAVs observed by \textit{Kepler}/\textit{K2} at different temperatures across the instability strip \citep{2017ApJS..232...23H}. \\
The top panels shows where DA white dwarfs pulsate in the spectroscopic parameter space of log-surface-gravity ($\log{g}$) and effective temperature. Circle and square markers indicate the parameters of some pulsating DAV stars. Small crosses show white dwarfs not observed to vary, but within the instability strip this is likely due to insufficient signal-to-noise of the data. Dash-dotted lines show how models of white dwarfs of masses 0.6\,$M_\odot$ and 0.8\,$M_\odot$ cool across the parameter space. The blue and red edges of the instability strip are marked with dashed lines. Red circles indicate outbursting DAVs (Section~\ref{sec:oDAVs}).\\
The bottom panels display the periodograms of the \textit{Kepler}/\textit{K2} light curves of, from top to bottom, five progressively cooler DAVs. Their locations in the instability strip are indicated with square markers of corresponding color and letter in the upper plot. As DAVs cool across the pulsational instability strip, they tend to drive pulsations with progressively longer periods (lower frequencies). The higher-frequency modes appear as sharp features in the periodogram, indicating a high degree of coherence, while low frequency modes ($\lesssim 1250\,\mu$Hz) are less coherent, with power smeared across sampled frequencies (Section~\ref{sec:coherence}). The highest amplitudes are observed near the middle of the instability strip. \\
\textcopyright\ AAS. Reproduced from \citet{2017ApJS..232...23H} with permission.
}
\label{fig:davstrip}
\end{figure}

\subsubsection{DAVs: hydrogen-atmosphere pulsating white dwarfs (ZZ Ceti stars)}

Hydrogen-atmosphere white dwarf stars, known as DA white dwarfs, are observed to pulsate when their effective temperatures fall between 10{,}500--13{,}000\,K \citep{2017ApJS..232...23H}. The pulsating DAs are called DAVs (``V'' for ``variable''), or ZZ Ceti stars after one of the earliest DAVs discovered \citep{1971ApJ...163L..89L}. The hot and cool limits of the instability strip are referred to as the blue and red edges. As a DA white dwarf cools to the temperature of the blue edge, electrons in the atmosphere are able to recombine with hydrogen nuclei, increasing the opacity and impeding the outward flow of radiation sufficiently to establish an outer convection zone, which deepens as the star cools further. The pulsations are understood to be driven to observable amplitude by the convective driving mechanism, where the convection zone absorbs and releases energy in response to local radiative flux variations in a way that amplifies oscillations \citep{1991MNRAS.251..673B, 1999ApJ...511..904G}.

DAVs are the most populous class of pulsating white dwarf, with around 500 known examples \citep{2022MNRAS.511.1574R}. The boundaries of the instability strip are mass dependent, with more massive white dwarfs pulsating at higher temperatures \citep{2023MNRAS.522.2181K}, and low-mass white dwarfs pulsating at cooler temperatures \citep{2013MNRAS.436.3573H,2013ApJ...762...57V}. A sample of DAVs, and the blue and red edges of the instability strip, are shown in the spectroscopic parameter space of effective temperature and log surface gravity in the top panel of Fig.~\ref{fig:davstrip} (where higher surface gravities correspond to more massive white dwarfs).

With a large sample of DAVs available for study, observational trends across the instability strip are well established \citep{2006ApJ...640..956M,2017ApJS..232...23H}. 
Observations support that the DAV instability strip is ``pure,'' in the sense that all white dwarfs with hydrogen atmospheres are expected to pulsate as they cool through this region of parameter space \citep{2004ApJ...600..404B,2007A&A...462..989C}; therefore asteroseismic inferences about the structures of DAVs are representative of DA white dwarfs generally.
As DA white dwarfs cool across the instability strip, they are observed to pulsate with progressively longer periods on average. As the depth of the convection zone increases as a white dwarf cools, its thermal timescale increases, and the convective driving mechanism can more efficiently drive modes with longer periods \citep{1999ApJ...519..783W}. Pulsators near the middle of the strip are observed to exhibit the highest-amplitude variability. These trends are visible in the lower panels of Fig.~\ref{fig:davstrip}, which compare periodograms of progressively cooler DAVs. Distinctions in the relative coherence of modes with different pulsation timescales are described in \ref{sec:coherence}, and the outburst behavior observed near the red edge of the instability strip is detailed in \ref{sec:oDAVs}.

\subsubsection{DBVs: helium-atmosphere pulsating white dwarfs (V\,777 Her stars)}
Helium-atmosphere white dwarf stars are called DB white dwarfs, and the variables are DBVs. These stars are observed to pulsate when their effective temperatures fall between 22{,}000--31{,}000\,K \citep{2022ApJ...927..158V}. DBVs are sometimes called V\,777 Her stars after their prototype \citep{1982ApJ...262L..11W}. By analogy to DAVs, which pulsate at effective temperatures where a partial ionization of hydrogen first establishes an outer convection zone, the DBs were theoretically predicted to pulsate at higher temperatures where the helium atmosphere becomes partially ionized \citep{1982ApJ...252L..65W}. This allows convection to drive pulsations as it does for DAVs \citep{2017ASPC..509..321V}. There are around 50 DBVs known to date \citep{2022A&A...668A.161C}, and their observed mode characteristics across the DBV instability strip generally mimic the behaviors of DAVs \citep{2022ApJ...927..158V}. 

\subsubsection{GW Vir stars: hot pulsating white dwarfs and pre-white dwarfs} 

Following the ejection of their outer layers as planetary nebulae, the remnant cores of dying stars contract onto the white dwarf cooling track as PG\,1159 stars. 
Many of these are observed to pulsate as ``GW Vir'' stars with effective temperatures in the range 80{,}000--180{,}000\,K \citep{2023ApJS..269...32S}. The GW Vir instability strip has been referred to as the ``DOV'' instability strip, and some pulsating stars still enshrouded in planetary nebulae have been called PNNVs, for ``planetary nebula nucleus variables.''  Stars cross this instability regime twice, once as they are contracting and increasing in effective temperature, and then again as they begin cooling at the hot end of the white dwarf cooling track. Unlike the DAV and DBV classes, not all stars in the GW Vir instability strip are observed to pulsate; only around 1/3 of stars in the mass--$T_\mathrm{eff}$ space of the GW Vir instability strip pulsate \citep{2023ApJS..269...32S}, and those with more nitrogen in their atmospheres are more likely to show detectable variability \citep{2021ApJ...918L...1S}. These hot white dwarfs do not have outer convection zones, and their pulsations are driven by the kappa-gamma mechanism operating on the K-shell electrons of carbon and oxygen \citep{1996ASPC...96..361S,2005A&A...438.1013G}. There are two dozen GW Vir pulsators known to date \citep{2022MNRAS.513.2285U}.

\subsection{Nonradial gravity mode oscillations}\label{chap2:subsec1}

Each resonant frequency of the star corresponds to a different mode of oscillation, and pulsating white dwarfs typically oscillate in many modes simultaneously.
The pulsations observed from white dwarf stars are theoretically and observationally \citep{1982ApJ...259..219R} consistent with gravity mode oscillations (g-modes), where displacements within the star are restored by buoyancy.  As opposed to pressure modes (p-modes), which are the other most common type of stellar pulsation observed in other types of star (including the Sun), gravity modes have longer periods and involve mostly horizontal displacements \citep{2010aste.book.....A}.  
Gravity modes have specific characteristics that can guide the interpretation of detected pulsation signals (\ref{sec:mps}).

\begin{figure}[t]
\centering
\includegraphics[width=.7\textwidth]{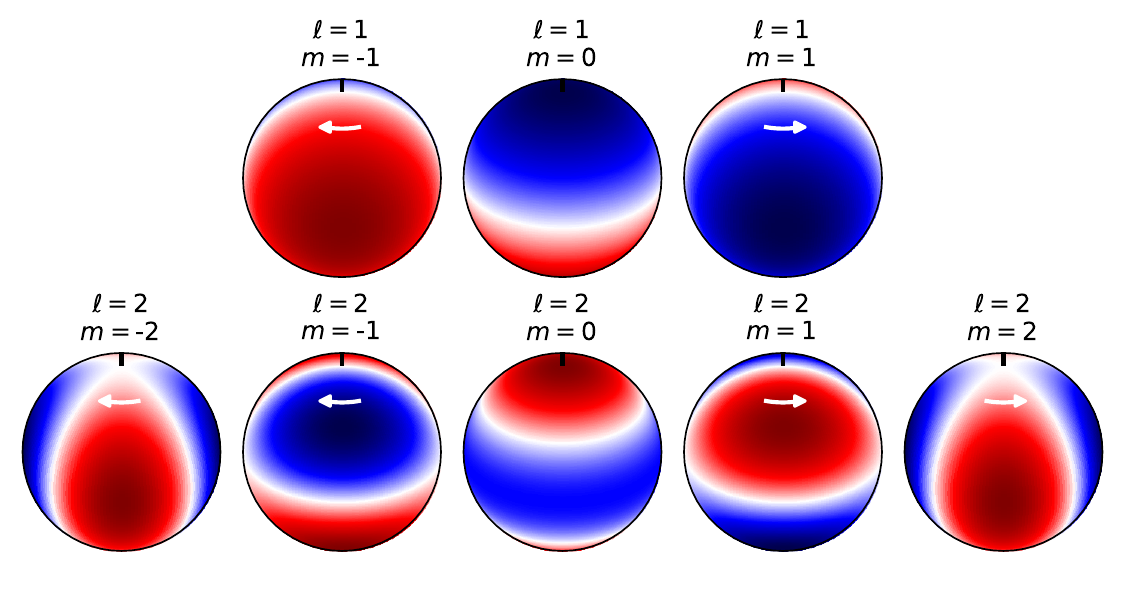}
\caption{Spherical harmonic patterns of stellar pulsation for $\ell=1$ and 2 modes that are observed from pulsating white dwarfs. The $m=0$ modes are standing waves on the surface that oscillate in place. The $m\neq0$ modes are traveling waves moving in opposite directions on the surface depending on the sign of $m$, as indicated with the white arrows. Maps are displayed with a 60 degree inclination, and the rotational poles are marked with black lines.}
\label{fig:sphericalharmonics}
\end{figure}

The discrete set of pulsation modes of a spherical star are three dimensional, and the oscillations are characterized with three quantum numbers: $k$, the radial overtone number (called $n$ in some areas of asteroseismology); $\ell$, the spherical degree; and $m$, the azimuthal order. 
All g-modes are non-radial oscillations, and every mode has an angular dependence that takes the pattern of spherical harmonics for a spherical star (\ref{sec:sph}), which are described by $\ell$ and $m$. The number $k$ describes the radial dependence, and the oscillation frequencies that will be resonant within a star can be best understood from propagation diagrams (\ref{sec:prop}).

\subsubsection{Spherical harmonics}\label{sec:sph}

The patterns of pulsations at the stellar surface are spherical harmonics. Spherical harmonics are identified by two quantum numbers: the spherical degree, $\ell$, says how many nodal lines are on the surface, and the azimuthal order, $m$, says how many of these pass through the (rotational) poles of the star. The spherical harmonics for $\ell=1$ and 2 are shown in Fig.~\ref{fig:sphericalharmonics}. White dwarfs typically pulsate in many modes simultaneously, each taking the form of a spherical harmonic oscillating with its own frequency and amplitude.

\paragraph{Spherical degree ($\ell$)}
For nonradial oscillations, $\ell$ is an integer of one or greater. For higher $\ell$, the surface of the star gets sliced into increasingly many small sectors, and for each patch of the surface that is getting brighter during a pulsation cycle, there is a neighboring region that is getting darker. When observing the disc-integrated light from a star, adjacent regions tend to cancel each other out when there are a large number of nodal lines on the surface \citep{1977AcA....27..203D}, and pulsation signals observed from white dwarfs are usually assumed to be $\ell=1$ or $\ell=2$ modes.

\paragraph{Azimuthal order ($m$)}
The number of the $\ell$ nodal lines on the stellar surface that pass through the poles is recorded as $m$. For white dwarfs, the poles are typically thought to correspond to the rotation axis. If $m=0$, all $\ell$ nodal lines slice through the star parallel to the equatorial plane. This separates the surface into neighboring regions that vary in brightness as a standing wave, each in anti-phase to its neighbors. If $m\neq0$, the patterns shown in Fig.~\ref{fig:sphericalharmonics} appear as traveling waves on the surface. Whether $m$ is positive or negative determines whether the spherical harmonics travel in the prograde or retrograde direction with respect to rotation, and it must be the case that $|m|\leq\ell$.

\subsubsection{Propagation diagrams}\label{sec:prop}

\begin{figure}[t]
\centering
\includegraphics[width=.6\textwidth]{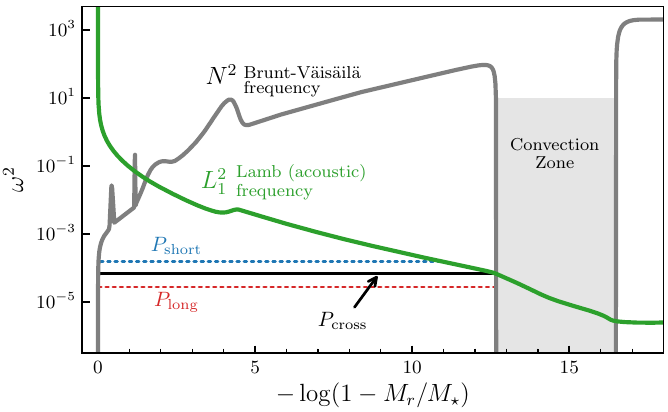}
\caption{A propagation diagram from a stellar model of a DAV pulsator. See section~\ref{sec:prop} for discussion. \textcopyright\ AAS. Reproduced from \citet{2020ApJ...890...11M} with permission.}
\label{fig:prop}
\end{figure}

To understand the radial dependence of white dwarf pulsations, it is useful to consider propagation diagrams like the one displayed in Fig.~\ref{fig:prop}. Propagation diagrams are a representation of the physical properties of a stellar model that are relevant to tuning the pulsation frequencies. They show how the characteristic Brunt-V{\"a}is{\"a}il{\"a} (BV) and Lamb frequencies vary through the stellar model from center (left) to surface (right). The BV frequency is the frequency at which a displaced portion of the star would oscillate about its equilibrium position at different depths within the star due to buoyancy, and the Lamb frequency is a characteristic acoustic frequency (with dependence on $\ell$).  These are commonly plotted against mass coordinate for white dwarfs written as $-\log{(1-M_r/M_\star)}$, where a number $X$ on the x-axis means that $10^{-X}$ of the star by mass is to the right of that point on the diagram; this makes details of the thin white dwarf envelope discernible in graphs.

Regions of transition in the properties of the stellar model impart signatures in the characteristic frequencies. In Fig.~\ref{fig:prop}, local bumps in the BV frequency can be identified to correspond to the base of the hydrogen envelope ($-\log{(1-M_r/M_\star)} = 4$), the base of the helium layer ($=2$), and at steep gradients in the carbon/oxygen ratio within the core. Gravity waves cannot propagate into a convective region, so the BV frequency drops sharply to zero at the base of the convection zone.

The gravity modes observed in white dwarfs oscillate in regions of the star below both the BV and Lamb frequencies. No pressure mode pulsations have been definitively detected from a white dwarf. The resonant frequencies of the gravity modes are tuned such that the oscillations just fit in the cavity below the BV and Lamb frequencies, with $k$ nodes in between. Modes with higher radial overtone number, $k$, oscillate with longer periods (lower in the propagation diagram). The resonant pulsation frequencies of white dwarfs are acutely sensitive to local features in the BV and Lamb frequencies, and therefore provide a powerful diagnostic for inferring the details of white dwarf interior structures.

\subsubsection{Mean period spacing and deviations}\label{sec:mps}

A useful property of gravity modes is that, in the asymptotic limit of high radial overtone number, $k$, successive radial overtones are evenly spaced in period \citep{2010aste.book.....A}. That statement applies to axisymmetric ($m=0$) modes of the same $\ell$ value. 
For $k \gg \ell$, the spacing between $m=0$ pulsation periods of subsequent $k$ is, on average, 
\begin{equation}
    \Delta\Pi^a_\ell = \frac{\Pi_0}{\sqrt{\ell(\ell+1)}},
\end{equation}
where $\Pi_0$ is a constant.

The behavior in the asymptotic limit is an idealization, and real white dwarf pulsation modes are observed (and calculated) to roughly align with an even period spacing on average (for $k>>\ell$). The ``mean period spacing'' is most sensitive to the global stellar parameters of mass and effective temperature, though other parameters also have some influence (H layer mass in DAVs, especially; \citealt{1990ApJS...72..335T}). Identifying the pattern of average spacing in a set of measured pulsation periods can constrain these global properties, and can guide the further interpretation of the individual modes. A few statistical techniques commonly used to identify potential mean period spacings are the Kolmogorov–Smirnov \citep{Kawaler88}, Inverse Variance \citep{ODonoghue94}, and Fourier Transform \citep{Winget91} tests.

The deviations from a perfectly even period spacing carry information about aspects of the stellar interior structure that cause localized features in the BV and Lamb frequencies. A useful analogy can be made to the pattern of standing waves on a string \citep{2005ASPC..334..553M}. For a string of uniform density, the frequencies of successive overtones are separated by an equal amount, in the same way that g-modes in a smooth, idealized star have an even period spacing. However, if the string is not uniform, and has one or more localized regions where the wave speed changes, the pattern of overtones will depart from the even spacing in frequency, though the deviations will vary about an even average spacing. The same happens in a white dwarf due to bumps in the BV frequency. Any localized feature will modify the resonant wave functions by an amount that depends on its location relative to the radial nodes. The deviations about the mean period spacing are cyclic, and the pattern of deviations is known as the mode trapping cycle. The specifics of these deviations are decided by the specifics of the localized features in the propagation diagram, and therefore the individual mode periods carry information about the precise interior structure profile of a pulsating white dwarf star \citep{2002A&A...387..531C}.

\subsubsection{Rotational splittings}

\begin{figure}[t]
\centering
\includegraphics[width=.95\textwidth]{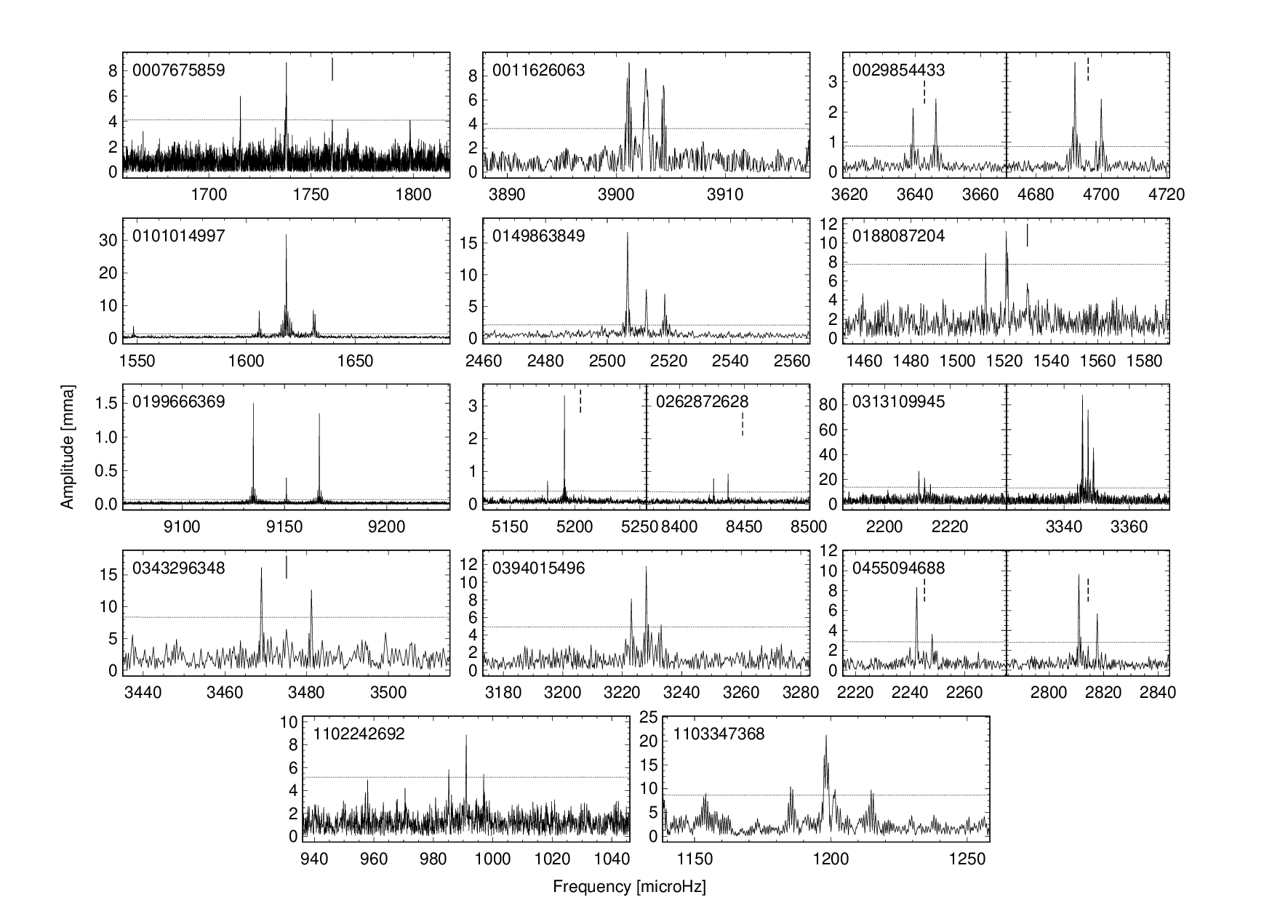}
\caption{Examples of rotational splittings of $\ell=1$ modes in DAV pulsators observed with TESS, from \citet[reproduced with permission from the authors]{2024A&A...684A..76B}. Dotted horizontal lines shows the significance threshold used in that work to identify signals that are significantly above the noise. Solid vertical markers show where candidate components of the triplets are below the significance threshold, and dotted vertical lines show periods measured from previous observations.}
\label{fig:splittings}
\end{figure}

In a spherically symmetric star, the spherical harmonics of different $m$ are degenerate. Rotation of a white dwarf can lift this degeneracy, splitting modes of each $\ell$ and $k$ into ($2\ell+1$) components with different frequencies. In the slow-rotating limit, these rotationally split modes are separated in frequency from the $m=0$ mode by an amount
\begin{equation}
    \delta\nu_{k,\ell,m} = m(1-C_{k,\ell})\Omega
\end{equation}
where $\Omega$ is the rotation frequency, and $C_{k,\ell}$ are the Ledoux constants that describe the effect of the Coriolis force on the modes \citep{1951AnAp...14..438L}. For gravity modes, $C_{k,\ell}$ asymptotically approaches $1/(\ell(\ell+1))$ in the limit of high $k$, and it is generally less than this value below the asymptotic limit.

Identification of rotational multiplets in the periodograms of white dwarfs can reveal white dwarf interior rotation rates. \citet{2017ApJS..232...23H} found that more massive white dwarfs have generally faster rotation rates, evidenced by the larger splittings of rotational multiplets. Figure~\ref{fig:splittings} shows examples of rotational $\ell=1$ triplets identified in periodograms of TESS light curves from \citet{2024A&A...684A..76B}.

Rotational splittings also aid in mode identification, as $\ell=1$ modes are split into triplets, and $\ell=2$ modes are split into quintuplets with a different frequency spacing. However, it is not common for all components of rotational multiplets to be driven to observable amplitudes in white dwarfs, and most observed modes are not clearly part of multiplets. In the absence of observing sufficient multiplet components to make clear $m$ identifications, asteroseismic analyses should avoid assumptions about the $m$ value of observed modes.

\subsection{Nonlinear pulsations}\label{sec:nonlinear}

The structure of the white dwarf can adjust dynamically to the stellar pulsations, and since the pulsations all interact with the star, nonlinear effects can arise. In DAVs and DBVs, the pulsations are driven by the convection zone, and the convection zone adjusts on a timescale much faster than the pulsation timescale. As a result, the convection zone varies in depth in response to local variations in temperature. The deeper the convection zone, the more the pulsation signals are attenuated at the photosphere. During a pulsation cycle, the outflowing energy reaches the photosphere more readily through hot regions of the star, where the convection zone is locally thinner. This causes the photometric variations due to pulsations to not be strictly sinusoidal, but to have sharper peaks at maximum brightness \citep{1992MNRAS.259..519B}.

Modeling the pulsation/convection interaction can aid with mode identification and provides an avenue to study the physics of stellar convection. For a pulsation that is sinusoidal at the base of the convection zone, \citet{2005ApJ...633.1142M} shows how modeling the absorption of energy by the convection zone can produce non-sinusoidal flux variations at the photosphere that match observations. The precise shape of pulsation signals in the light curve is determined by how sensitively the depth of the convection zone responds to the pulsations. \citet{2005ApJ...633.1142M} parameterize this sensitivity in terms of the thermal response timescale of the convection zone, $\tau_C$,
\begin{equation}
    \tau_C = \tau_0 \left(\frac{T_{\rm eff}}{T_{\rm eff,0}}\right)^{-N}
\end{equation}
where $T_{\rm eff}$ is the local effective temperature that varies due to pulsations, and $T_{\rm eff,0}$ and $\tau_0$ are average effective temperatures and thermal response timescales of the convection zone. The exponent $N$ captures how rapidly the convection zone changes in size due to local temperature variations; mixing length theory and fits to light curves say that the response is highly sensitive, with typical values as large as $N\approx90$ and $\approx23$ in the DAV and DBV instability strips, respectively \citep{2005ApJ...633.1142M}. The results are dependent on viewing angle (inclination) and the spherical harmonic pattern observed ($\ell$ and $m$), which nonlinear pulse shapes can also constrain.

Fourier-transform-based periodograms decompose light curves into sinusoids, and nonsinusoidal pulsations will show up as a series of harmonics. The harmonics are at exact integer multiples of the pulsation frequency and combine to reproduce the pulse shape (as a Fourier series). For multiperiodic pulsators, there are additional combination frequencies that appear at precise sums and differences between independent pulsation frequencies. It is essential to white dwarf asteroseismology to identify the periodogram peaks associated with combination frequencies and harmonics, so as not to interpret them as independent resonant frequencies of the star. Once identified, the amplitudes and phases of combination frequencies can aid in mode identification \citep{2001MNRAS.323..248W,2005ApJ...635.1239Y}.

\subsection{Mode coherence}\label{sec:coherence}

Some white dwarfs appear to pulsate in an incredibly coherent fashion, with pulsation signals phasing up over years of observations \citep{1982ApJ...254..676K}. The pulsation modes of other white dwarfs can change dramatically in phase and amplitude on timescales of days, often looking like completely different pulsators from one observing run to the next \citep{1998ApJ...495..424K}. The distinction between these behaviors has reinforced an appreciation for the interactions between pulsations and the convection zone, while the most coherent signals provide opportunities to probe the extreme physics of white dwarfs and the presence of orbiting planets.

\subsubsection{A dichotomy of coherence}

Before the advent of precision survey photometry from space missions like \textit{Kepler} and TESS, a distinction was noted in the behavior of pulsating white dwarf stars. In DAV pulsators, this difference in behavior was initially interpreted as a temperature effect: hot DAVs near the blue edge of the instability strip tend to have coherent pulsations, and cool DAVs near the red edge are incoherent pulsators \citep{2006ApJ...640..956M}.

Continuous high-cadence monitoring from space has clarified the picture. In the periodogram of a continuous time series, the signature of a coherent signal is one sharp, dominant peak. On the other hand, signals that are not coherent on timescales shorter than the span of observations will appear as wide bands of power distributed around a central frequency. These incoherent modes look similar to solar-like oscillations, which are stochastically driven and damped by convection, and they can be fit with Lorentzian functions in the power spectrum as is commonly done for solar-like oscillators \citep{2015ApJ...809...14B}.

The work of \citet{2017ApJS..232...23H} found that higher frequency pulsations in white dwarfs tend to be coherent, while lower frequency modes are typically incoherent. This can be seen in the bottom panels of Fig.~\ref{fig:davstrip}, where pulsations with periods shorter than $\approx800$\,s (frequency $\gtrsim$ 1250\,$\mu$Hz) appear as sharp peaks, while power from modes with longer periods is smeared through a small range of sampled frequencies. While the temperature of the star determines which modes are most likely to be excited, both coherent and incoherent modes can coexist within the same star (e.g., panel d of Fig.~\ref{fig:davstrip}).

\citet{2020ApJ...890...11M} provided a physical interpretation of the $\approx800$\,s timescale that separates the pulsation mode behavior into a dichotomy of coherence. In the propagation diagram shown in Fig.~\ref{fig:prop}, this timescale is labeled as $P_\mathrm{cross}$, where the BV frequency crosses the Lamb frequency at the base of the convection zone. Pulsation modes with periods shorter than $P_\mathrm{cross}$ are bounded by the Lamb frequency at their outer turning point, while longer-period modes propagate to the BV frequency that is defined by the base of the convection zone. As described in \ref{sec:nonlinear}, the base of the convection zone in a pulsating white dwarf varies in depth dramatically in response to local temperature variations from the pulsations, causing the size of the resonant cavity to vary. The long-period modes are therefore reflecting from a continuously moving boundary, likely disrupting the coherence of these modes.

\subsubsection{Monitoring coherent modes}

\begin{figure}[t]
\centering
\includegraphics[width=.45\textwidth,angle=-90]{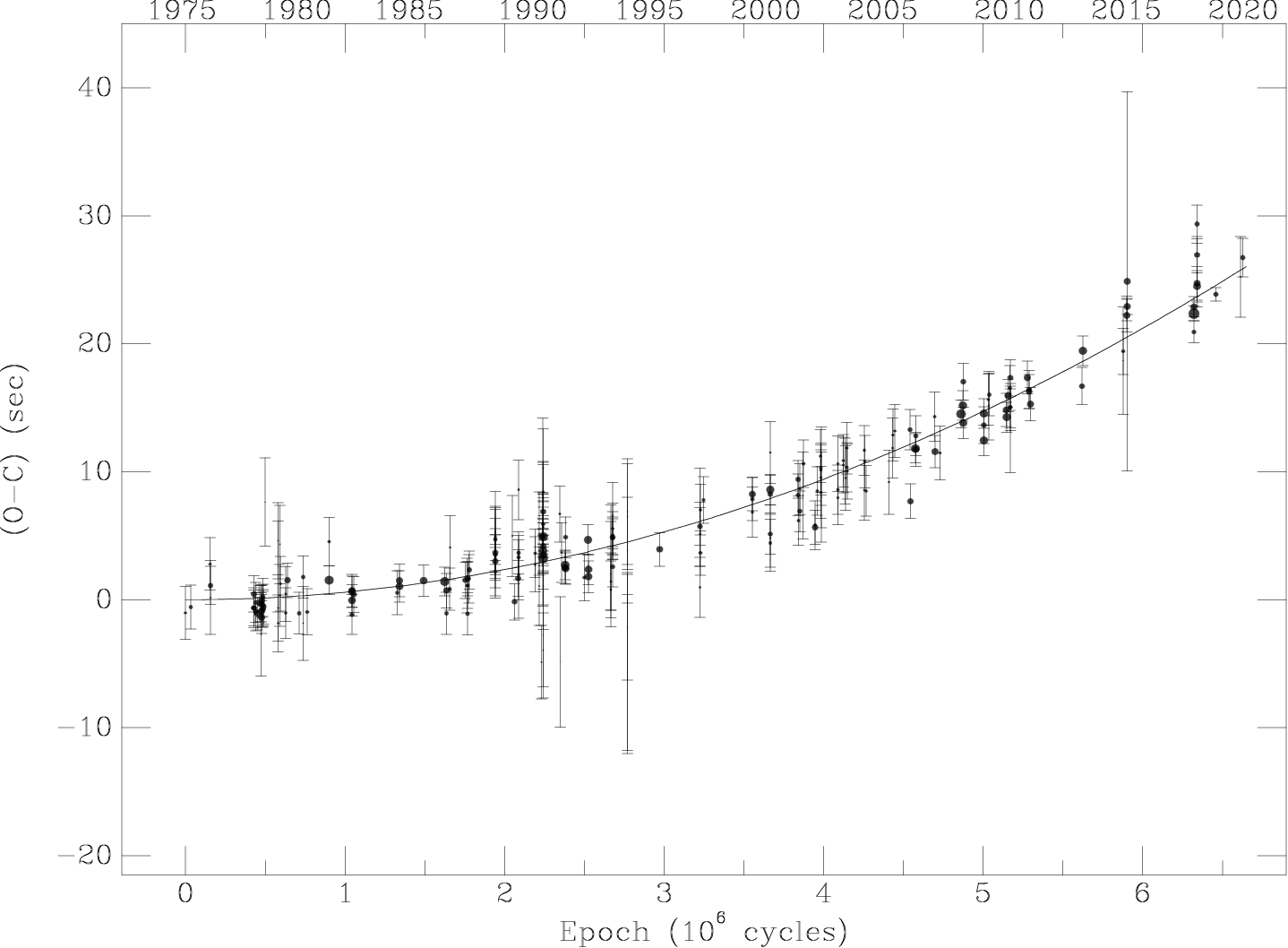}
\caption{The observed-minus-calculated ($O-C$) diagram for the 215-second mode of the DAV pulsator G\,117-B15A shows a gradual drift corresponding to a rate of period change of $(5.12 \pm 0.82) \times 10^{-15}$\,s\,s$^{-1}$. The top axis shows year of observation, and the bottom axis shows the number of pulsation cycles elapsed ($\times10^6$). \textcopyright\ AAS. Reproduced from \citet{2021ApJ...906....7K} with permission.}
\label{fig:ominusc}
\end{figure}

Unlike the low-frequency modes that are constantly jostled by the convection zone, high frequency pulsations propagate rather undisturbed, deeper within the star. These modes are exceedingly coherent, and some have been subject to decades-long monitoring to track gradual drifts of their pulsation signals. Over 45 years of observations of the 215-second mode in G\,117-B15A support that this pulsating white dwarf is the most stable optical clock known \citep{2021ApJ...906....7K}.

Even the most coherent white dwarf pulsations are not expected to be perfectly steady. White dwarfs evolve by cooling, and the subtle changes to their interior structures while they cool within the pulsational instability strips causes their resonant frequencies to drift slowly. Measuring the rate of pulsation period change ($\dot{P}$) sensitively tracks the rate of this secular cooling process. Measurements of white dwarf cooling rates can constrain exotic physics that could affect this process, such as the rate of change of the gravitational constant $G$, the magnetic moments of neutrinos, and properties of WIMPs and axions---both candidate dark matter particles \citep{2022FrASS...9....6I}.

Rates of pulsation period change are typically measured by tracking phase variations of the pulsation signals. An observed-minus-calculated ($O-C$) diagram shows the difference between measured arrival times of pulse maximum and the expected arrival times calculated with the assumption that the signal is strictly periodic \citep[Ch.\ 9.6]{robinson2016data}. The $O-C$ diagram for the DAV pulsator G\,117-B15A from \citet{2021ApJ...906....7K} is displayed in Fig.~\ref{fig:ominusc}. Relative to the signal phase when observations began in 1974, observed brightness maxima from pulsation of the 215-s mode have been gradually delayed, by only as much as 26 seconds over 45 years of observations. Because changes compound over time, the signature of a constant rate of period change in an $O-C$ diagram is a parabola. The measured rate of period change of this pulsation mode is $(5.12 \pm 0.82) \times 10^{-15}$\,s\,s$^{-1}$ \citep{2021ApJ...906....7K}, with similar rates measured for other DAVs \citep{2013ApJ...771...17M,2015ASPC..493..199S}, and faster rates of cooling detected for hotter DBV \citep{2011MNRAS.415.1220R} and GW Vir \citep{2008A&A...489.1225C} pulsators.

Another useful aspect of the coherence of high-frequency pulsations in white dwarfs is their potential to reveal orbital companions. If a white dwarf is orbited by a planet, the white dwarf will exhibit a reflex motion about the center of mass of the system. As this reflex motion modulates the distance to the white dwarf along our line of sight, pulsation signals will have to travel different amounts of time to reach our telescopes. Variations in light travel time from a white dwarf experiencing reflex motion produces phase variations that would appear sinusoidal (for a circular orbit) in an $O-C$ diagram. Since most stars end their lives as white dwarfs \citep{2008ApJ...676..594K}, and most stars host exoplanets \citep{2020AJ....159..248K}, we expect that most white dwarfs host exoplanets \citep{2015MNRAS.447.1049V,2021ARep...65..246A}. So far, attempts to detect exoplanets orbiting white dwarfs from phase modulations of pulsation signals has only yielded upper limits \citep{mullally,winget}.

\begin{figure}[t]
\centering
\includegraphics[width=.8\textwidth]{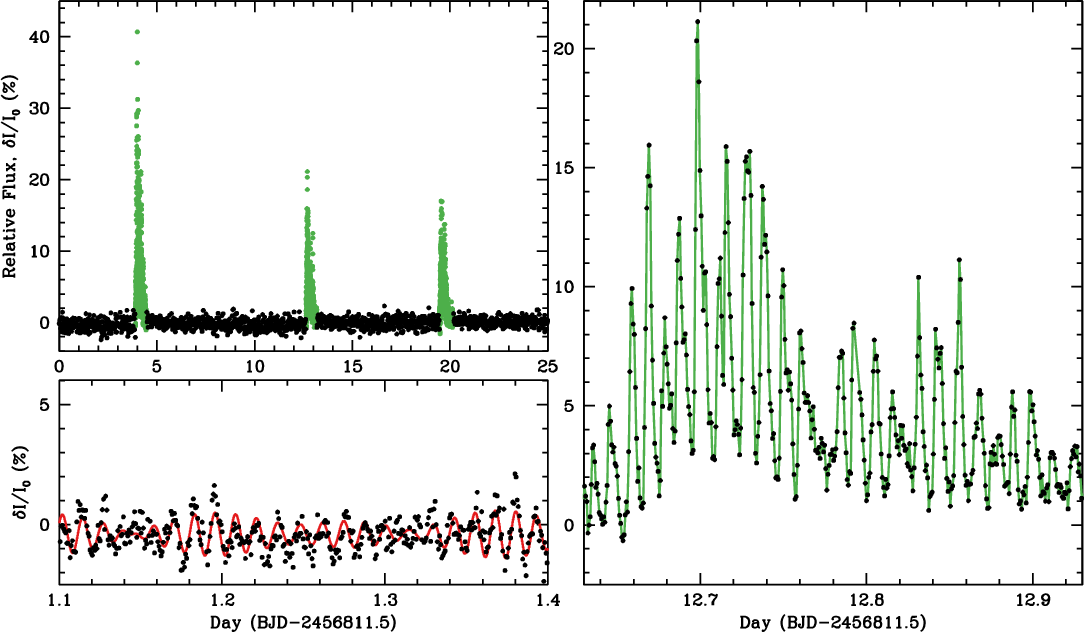}
\caption{Portions of the \textit{K2} light curve of the DAV pulsator PG 1149+057. The bottom left panel zooms in on a region of the light curve in quiescence, and the right panel reveals the detailed changes that occur during one outburst. Figure reproduced from \citet{2015ApJ...810L...5H} with permission.}
\label{fig:oDAV}
\end{figure}

\subsection{Pulsational outbursts}\label{sec:oDAVs}

Some pulsating white dwarf stars have been observed to show dramatic brightness increases that recur on timescales of days and last for many hours. This behavior was first identified in nearly continuous coverage of a DAV from \textit{Kepler} \citep{2015ApJ...809...14B}. These ``pulsational outbursts'' have only been observed in pulsating DA white dwarfs near the cool edge of the DAV instability strip (some are marked with red circles in Fig.~\ref{fig:davstrip}), where the behavior appears to be fairly common \citep{2016ApJ...829...82B}. In hindsight, the ``whoopsie'' or ``sforzando'' observed in 1996 from the DBV star GD\,356 \citep{2003A&A...401..639K,2009ApJ...693..564P} was likely the first recorded example of a pulsational outburst from a white dwarf. Fig.~\ref{fig:oDAV} shows portions of the \textit{K2} light curve of the outbursting DAV PG\,1149+057 \citep[from][]{2015ApJ...810L...5H}. During the outbursts, which can raise the mean brightness of the white dwarf by tens of percent, the pulsations also show an increase in observable amplitude.

An understanding is emerging about the mechanism that causes these outbursts. Somehow, the pulsational energy that builds in the self-driven pulsation modes is rapidly deposited into the outer layers of the star. It has been proposed that a driven pulsation couples to two other, globally damped modes through a parametric resonance; when the amplitude of the driven mode exceeds some threshold, it triggers a transfer of pulsational energy to the globally damped ``daughter'' modes \citep{2018ApJ...863...82L}. The temperature rises at the photosphere, causing an overall brightness increase as the star radiates away this excess energy. The convection zone responds to the temperature increase by getting thinner, attenuating remaining pulsation modes by a lesser amount, causing them to temporarily have higher observable amplitudes. Returning to the quiescent state, the driven modes start to build back up in amplitude, and the cycle begins again.

\section{White dwarf rotation}\label{sec:rotation}

Some white dwarf stars are not uniformly bright across their photospheres.  Such stars can exhibit photometric variability as rotation brings different parts of the surface into view. This enables surface rotation rates to be measured, as can be done for some other types of star. Suspected causes of rotational photometric variation in white dwarfs include nonuniform accretion of metals \citep{2000ApJ...537..977D}, or magnetic field geometries that cause different amounts of local magnetic Zeeman line splitting at the photosphere \citep{2023MNRAS.520.6111H}.

Rotational photometric variations are not always simply sinusoidal, and a periodogram can reveal multiple harmonics of the rotation frequency to represent the rotation profile. The underlying rotation frequency does not always have the highest amplitude; if the star has two bright/dark regions on opposite sides of its rotating surface, its dominant photometric variation will be at double the rotation frequency. Depending on the signal-to-noise ratio of your observations, there may not be a significant periodogram signal at the true rotation rate, and only the second harmonic might be detected. This is a challenge for stellar rotation studies of many types of star, and it is good to be aware that the dominant periodogram peaks from rotational variability will sometimes correspond to half the rotation period.

\begin{figure}[t]
\centering
\includegraphics[width=.8\textwidth]{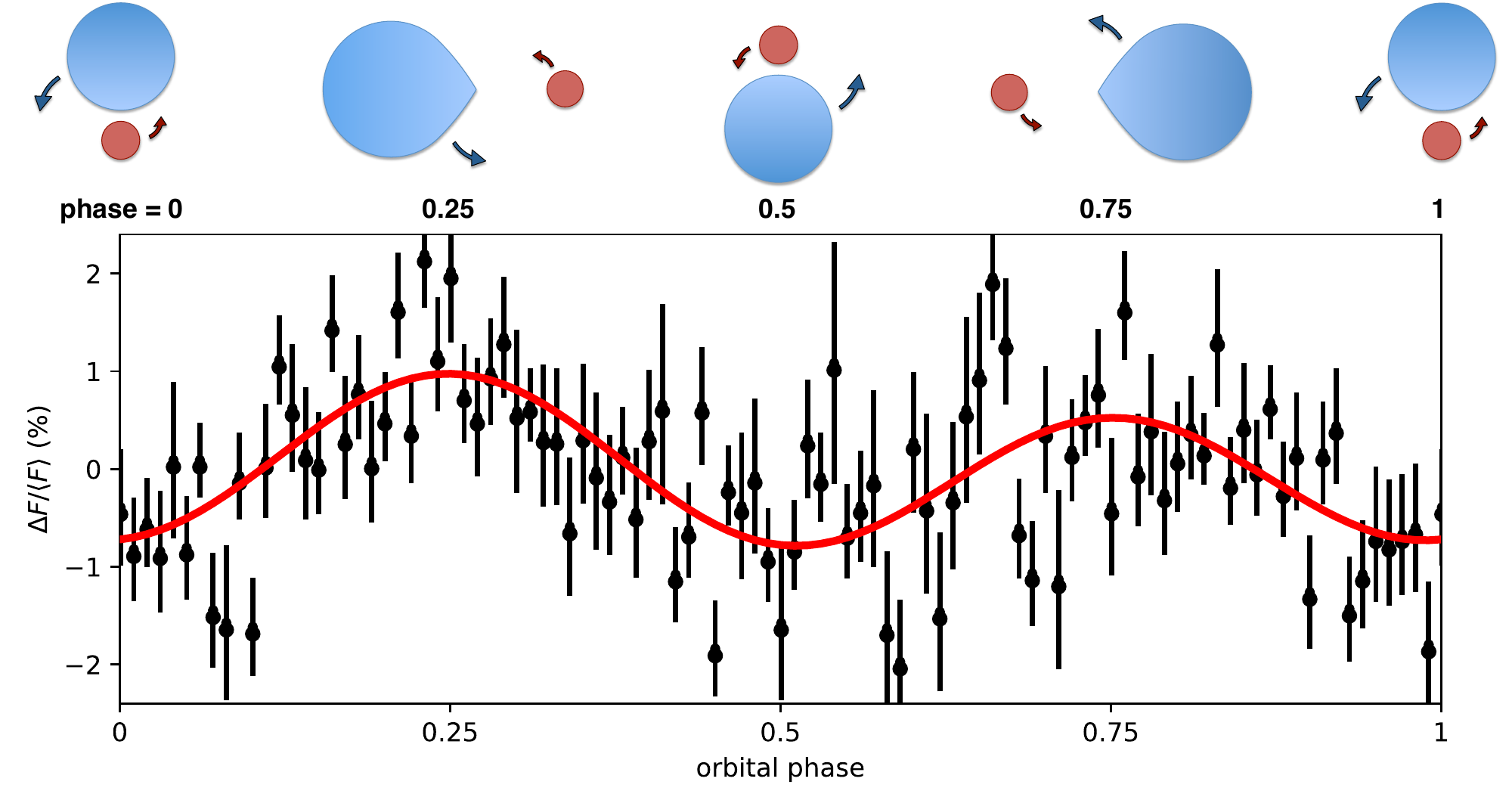}
\caption{A cartoon shows (not to scale) how tidal deformation of the extremely low-mass white dwarf primary of the binary system SDSS\,J1054$-$2121 corresponds to its photometric variability folded on the 2.5-hour orbital period. Reproduced from \citet{2018arXiv180905623B}.}
\label{fig:EVs}
\end{figure}

\section{Compact binaries}\label{sec:EVs}

Many white dwarf stars are found in binary systems, and there are a variety of ways that these systems can be photometrically variable. White dwarfs in binaries can eclipse and be eclipsed, there can be reflection effects, white dwarfs can gravitationally microlens their companions \citep{2014Sci...344..275K}, mass transfer onto white dwarfs through an accretion disk in cataclysmic variables can produce variability, this accreted material can ignite as novae, and enough mass transfer onto a white dwarf can overwhelm electron degeneracy pressure causing the star to explode as a Type Ia supernova. While all of great scientific interest, these processes are outside the scope of this chapter. We close by discussing one of the more common sources of photometric variability in binary systems: ellipsoidal variations from tidal distortion.

\subsection{Ellipsoidal variations in tight binary systems}

When two stars in a binary system are very close to each other, the tidal forces can be sufficient to distort the shape of one or both stars. The distortion of the star is aligned along the tidal axis connecting the two stars, and we get to view this distortion from different angles as the star moves through its orbit. Photometric variations are observed as the projected area of the distorted star changes over the orbital period. Figure~\ref{fig:EVs} shows a cartoon of how the changing viewing angle of a tidally distorted star produces the photometric ellipsoidal variations measured for the white dwarf binary SDSS\,J1054$-$2121 \citep{2018arXiv180905623B}. There are two maxima per orbital period (ellipsoidal variation signal appears in periodograms at twice the orbital frequency), since the broad side of the star is visible twice per orbit. One maximum is higher than the other because of Doppler beaming of light in the direction ahead of the primary star's motion \citep[the Doppler beaming signal appears at the orbital frequency;][]{2007ApJ...670.1326Z}. The ellipsoidal variations provide an opportunity to break the $sin{(i)}$ degeneracy from orbital inclination angle that limits our understanding of most non-eclipsing binary systems, and when analyzed with other data, they can constrain the mass--radius relationship for distorted white dwarfs.

White dwarfs are quite compact and therefore must be in very close binary systems for ellipsoidal variations to be significant. In fact, this effect is most often observed from ``extremely low mass'' (ELM; $\lesssim0.25\,M_\odot$) white dwarfs. From single-star evolution, white dwarfs of such low mass are expected to form from stars with main sequence lifetimes that exceed the age of the Milky Way \citep{2007ApJ...671..761K}. The ELM white dwarfs that we observe are expected to have formed through mass loss in tight binary systems, which is supported by the high fraction observed to still be in short-period binaries \citep{1995MNRAS.275..828M,2016ApJ...818..155B}. Since less massive white dwarfs have larger radii, the ELM white dwarfs are more deformable, and since most are in tight binary systems, many ELM white dwarfs exhibit ellipsoidal variations.
Analyzing the ellipsoidal variation signatures can constrain the properties of the ELM white dwarfs and their binary companions, and can provide insights into their evolutionary pathways through binary mass transfer \citep{2014ApJ...792...39H}.

\section{Conclusions}\label{sec:conc}

The photometric variability of white dwarf stars can provide valuable insights into the physical properties of these compact stellar remnants. Stellar pulsations in particular can be used to constrain white dwarf interior structures, especially those with many modes that propagate in different regions within these stars. Brightness variations from rotation or photometric binaries typically appear at rotation/orbital frequencies and/or harmonics of these frequencies; these scenarios can be difficult to tell apart in some cases from photometry alone, but close binaries will show radial velocity variations in spectroscopic data with a consistent periodicity.

Recent and planned astronomical surveys are increasingly collecting photometric data in the time domain, revealing an abundance of variable white dwarf systems. The ideal light curves for unambiguous characterization of variability are continuous, with a rapid frame rate, recorded over a long time baseline, such as those collected from the \textit{Kepler} and \textit{TESS} missions. Recent studies of white dwarfs have benefited tremendously from precision astrometry from \textit{Gaia} \citep{Gaia}, which has recently revealed hundreds of thousands of white dwarf stars with significant parallaxes \citep{2021MNRAS.508.3877G}. Most white dwarfs brighter than an apparent \textit{Gaia} magnitude of $G\approx 20$ have now been identified, informing efforts to discover and characterize variable white dwarfs.

\begin{ack}[Acknowledgments]

Thank you to the following authors for agreeing to allow their figures to be reproduced in this chapter: J.~J.~Hermes, M.~H.~Montgomery, Zs.~{Bogn{\'a}r}, and S.~O.~Kepler.

\end{ack}


\bibliographystyle{Harvard}
\bibliography{reference}

\end{document}